\documentclass[sigconf]{acmart}

\AtBeginDocument{%
  }

\usepackage{float}
\usepackage{hyperref}
\usepackage{enumitem}
\usepackage[noabbrev,capitalise]{cleveref}
\usepackage{xcolor}
\usepackage{balance}

\setlist[itemize]{topsep=2pt, itemsep=1pt, parsep=0pt, partopsep=0pt}
\setlist[enumerate]{topsep=2pt, partopsep=0pt, itemsep=1pt, parsep=0pt}

\copyrightyear{2026}
\acmYear{2026}
\setcopyright{cc}
\setcctype{by}
\acmConference[IDE '26]{3rd International Workshop on Integrated Development Environments }{April 12--18, 2026}{Rio de Janeiro, Brazil}
\acmBooktitle{3rd International Workshop on Integrated Development Environments (IDE '26), April 12--18, 2026, Rio de Janeiro, Brazil}
\acmPrice{}
\acmDOI{10.1145/3786151.3788605}
\acmISBN{979-8-4007-2384-1/2026/04}

\begin{document}

\title{In-IDE Toolkit for Developers of AI-Based Features}

\author{Yaroslav Sokolov}
\affiliation{%
  \institution{JetBrains}
  \city{Berlin}
  \country{Germany}
}
\email{yaroslav.sokolov@jetbrains.com}
\authornote{Both authors contributed equally to this research.}

\author{Yury Khudyakov}
\affiliation{%
  \institution{JetBrains}
  \city{Munich}
  \country{Germany}
}
\email{yury.khudyakov@jetbrains.com}
\authornotemark[1]

\author{Lenar Sharipov}
\affiliation{%
  \institution{JetBrains}
  \city{Munich}
  \country{Germany}
}
\email{lenar.sharipov@jetbrains.com}

\author{Andrei Gasparian}
\affiliation{%
  \institution{JetBrains}
  \city{Amsterdam}
  \country{Netherlands}
}
\email{andrei.gasparian@jetbrains.com}

\author{Parth Tiwary}
\affiliation{%
  \institution{JetBrains}
  \city{Krakow}
  \country{Poland}
}
\email{parth.tiwary@jetbrains.com}

\author{Artem Trofimov}
\affiliation{%
  \institution{JetBrains}
  \city{Berlin}
  \country{Germany}
}
\email{artem.trofimov@jetbrains.com}

\begin{abstract}
AI-enabled features built on LLMs and agentic workflows are difficult to test, debug, and reproduce, especially for product-focused software engineers without a machine learning background. We present the \emph{AI Toolkit}\footnote[1]{\url{https://plugins.jetbrains.com/plugin/26921-ai-agents-debugger/}} plugin for JetBrains IDEs, which brings tracing and evaluation directly into the Run/Debug loop. A mixed methods study with practitioners presents three consistent needs: (1) make evaluation regular and repeatable, (2) expose traces at the moment of execution, and (3) minimize setup and context switching. Guided by these needs, the AI Toolkit introduces an IDE-native workflow: run-triggered trace capture; immediate, hierarchical inspection; one-click “Add to Dataset” from traces; and unit-test–like evaluations with pluggable metrics. The first release in PyCharm shows promising early signals -- strong conversion when promoted at Run, sustained usage among those who capture traces, and low churn -- suggesting that IDE-native observability lowers activation energy and helps developers adopt disciplined practices. We detail the design and implementation of the AI Agents Debugger and AI Evaluation, report initial adoption telemetry, and outline next steps to broaden framework coverage and scale evaluations. Together, these results indicate that integrating AI observability and evaluation into everyday IDE workflows can make modern AI development accessible to non-ML specialists while preserving software-engineering practices.
\end{abstract}

\ccsdesc[500]{Software and its engineering~Integrated and visual development environments}
\ccsdesc[500]{Computing methodologies~Machine learning}
\ccsdesc[500]{Computing methodologies~Artificial intelligence}
\ccsdesc[500]{Computing methodologies~Natural language processing}

\keywords{Integrated Development Environment, Machine Learning, Artificial Intelligence, AI Agents, Natural Language Processing}

\maketitle

\section{Introduction}

AI-enabled features are increasingly built on top of large language models (LLMs) and agentic workflows that orchestrate prompting, retrieval, tools, and memory. While these systems accelerate development, they are often experienced as \emph{black boxes}: behavior depends on data, context, model versions, and non-deterministic sampling, making failures hard to localize and successes hard to reproduce. As teams scale AI features, they need tools for observing agent behavior, reasoning about changes, and verifying that improvements transfer from local tests to real projects.

A growing ecosystem addresses parts of this need, including LangSmith~\cite{langsmith}, Weights \& Biases~\cite{wandb,weave}, MLflow~\cite{mlflow}, and LangFuse~\cite{langfuse}. Collectively, these tools provide (i) \emph{tracing} to capture prompts, tool calls, and intermediate states; (ii) \emph{evaluation} to compare variants with human or automated metrics; and (iii) \emph{experiment tracking} to record configurations and outcomes over time. However, these capabilities are typically offered as separate services or dashboards, optimized for ML practitioners who already know which levers to pull and how to stitch them together.

We observe an emerging group of \emph{AI-based feature developers}: product-focused software engineers who add LLM/agent capabilities to existing applications without a background in ML. This cohort is growing rapidly -- Gartner projects that by 2027, 55\% of software engineering teams will build LLM-based features~\cite{gartner}. To solve day-to-day problems, these developers often adapt familiar monitoring tools rather than adopt new AI-native solutions~\cite{so2025-ai-observability-security}.

There is a clear gap between \emph{AI-native tooling} and \emph{AI-based feature developers}.
In our understanding there are several reasons for that: first, these developers lack the intuition that the new development approach is needed for AI systems, including (a) systematic \emph{evaluation} beyond ad hoc testing, (b) structured \emph{experiments} rather than one-off tweaks, and (c) \emph{observability} of the behavior of non-deterministic AI systems.
Second, the existing AI-native tooling is optimized for ML developers and is not intuitive for Software Engineers. Even when powerful tools are available, the user journey is unclear: developers are dropped into rich dashboards without guidance on 'what to do next'.
Crucially, \emph{AI-based feature developers} live in already well established ecosystem~\cite{so2025-tech-dev-envs-prof-ai} which includes IDEs, not in external web consoles. Setting up a context outside the coding environment lowers adoption of existing \emph{AI-native tooling}.

To bridge this gap we propose our AI Toolkit plugin\footnotemark[1] for JetBrains IDEs~\cite{pycharm}, bringing observability and evaluation workflows into the developer’s primary environment. Our contributions are:
\begin{itemize}
  \item \textbf{User needs research.} Using data from a developer survey conducted by the JetBrains Strategic Research \& Market Intelligence team together with our curated interviews, we characterize common pain points that we will address by our tooling.
  \item \textbf{Design and implementation.} We propose an IDE-native AI Toolkit which ties together tracing and evaluation aligned with everyday coding tasks. Then we describe the AI Toolkit plugin architecture and its integration points with project code, run configurations, and test artifacts, enabling low-friction setup and repeatable runs.
  \item \textbf{Release \& early signals.} We summarize initial adoption indicators and usage telemetry reflecting whether the IDE-native approach helps developers make evaluation and tracing regular parts of their workflow.
\end{itemize}

\noindent Taken together, our work aims to reduce the cognitive overhead of modern AI development by meeting developers where they work, guiding them from “I just want this agent to behave” to a disciplined loop of tracing and continuous evaluation inside IDE.

\section{Developer Needs \& Method}

To surface developer needs and implications for IDE-native observability and evaluation, we combined data from a developer survey conducted by the JetBrains Strategic Research \& Market Intelligence team with our own semi-structured interviews of product-focused engineers building LLM/agent features.

\subsection{Survey Insights}
The JetBrains Strategic Research \& Market Intelligence team recruited respondents through the JetBrains Research Panel in two groups: (i) AI/ML practitioners or employees of AI-active companies (as profiled in the Developer Ecosystem Survey, 2024), and (ii) a general coding audience matching the ecosystem survey on company size, language/IDE usage, and experience. Multi-select questions mean table percentages may exceed 100

\paragraph{Question: What are or were the main challenges you have faced while working on your AI-powered apps or features?}
\begin{center}
\footnotesize
\scalebox{1.0}{
\begin{tabular}{p{0.68\linewidth}r}
\toprule
Response option & Share \\
\midrule
Difficulty in testing AI features due to non-deterministic behavior & 56\% \\
High costs associated with AI services or infrastructure & 53\% \\
Data privacy or regulatory constraints & 52\% \\
Insufficient quality or accuracy of AI models & 50\% \\
Challenges in building robust evaluation pipelines & 36\% \\
Complexity or lack of AI-specific tools or frameworks & 28\% \\
Other & 1\% \\
None & 3\% \\
\bottomrule
\end{tabular}
}
\captionof{table}{Top challenges in AI feature development.}
\label{tab:challenges}
\end{center}

\noindent Testing under non-determinism (56\%) and model-quality concerns (50\%) remain difficult, and teams struggle to assemble robust evaluation pipelines (36\%) and AI-specific tooling (28\%). While high costs (53\%) and data privacy/regulatory issues (52\%) are substantial industry concerns, they are largely organizational/infrastructure problems beyond the scope of our IDE plugin; accordingly, we focus on tracing/observability and evaluation.

\paragraph{Question: Which features of the tools or platforms for developing AI agents did you or your company use?}

\begin{center}
\centering
\footnotesize
\scalebox{1.0}{
\begin{tabular}{p{0.68\linewidth}r}
\toprule
Response option & Share \\
\midrule
Standard built-in components (e.g., memory, planning) & 62\% \\
Tracing / Observability & 48\% \\
Evaluation of AI models' performance & 45\% \\
Advanced debugging (e.g., visual debugging) & 32\% \\
I don't know & 9\% \\
Other & 2\% \\
None & 4\% \\
\bottomrule
\end{tabular}
}
\captionof{table}{Adopted features in AI agent tooling.}
\label{tab:features}
\end{center}

\noindent While “built-in components” (62\%) dominate, they are framework-native and outside our integration scope. Apart from that, tracing/observability (48\%) and evaluation (45\%) rank as the most sought-after cross-framework capabilities, and interest in advanced debugging (32\%) indicates developers are actively seeking debugging support from tooling. This reinforces the need to surface these features directly in the IDE and to guide their effective use.

\paragraph{Question: How do you evaluate or test the performance of your AI-powered apps or features?}

\begin{center}
\footnotesize
\scalebox{1.0}{
\begin{tabular}{p{0.68\linewidth}r}
\toprule
Response option & Share \\
\midrule
I manually test AI features & 58\% \\
I use user feedback or production metrics & 42\% \\
We have dedicated team members (e.g., a team of ML engineers) & 32\% \\
I build automated evaluation pipelines using specialized tools & 23\% \\
We don’t have a formal evaluation approach in place & 10\% \\
\bottomrule
\end{tabular}
}
\captionof{table}{Evaluation and testing practices.}
\label{tab:eval-practices}
\end{center}

\noindent Developers rarely build automated evaluation pipelines; evaluation is predominantly manual or deferred to production feedback.

\paragraph{Implications.}
These findings validate our problem framing: teams lack clear, low-friction paths to systematic evaluation; an IDE-native workflow that unifies tracing and evaluation can meet developers where they work and convert ad hoc testing into repeatable practice.

\subsection{Interview Insight}
We conducted follow-up semi-structured interviews with eight engineers who build AI-powered applications. In each session, participants walked us through how they constructed their systems and how they determined when an agent’s output is “good enough,” with a focus on pre-release evaluation practices.

The core insight is that, without an ML background, pre-release evaluation is often absent: practitioners ship prototypes after manual self-checks and then depend on production metrics and user feedback to assess quality.
Elaborating on this, LLM-based development decouples evaluation from training, so teams frequently skip systematic evaluation entirely. Lacking ML intuitions (e.g., datasets, metrics, and experiment structure), they default to manual testing and in-production signals.

\subsection{Design Requirements}

Combining our mixed-methods findings, the survey highlights difficulty testing non-deterministic behavior (Table~\ref{tab:challenges}), strong demand for tracing/observability and evaluation (Table~\ref{tab:features}), and a reliance on manual testing or production feedback over automated pipelines (Table~\ref{tab:eval-practices}). Interviews further show that non-ML engineers often skip pre-release evaluation altogether. We therefore articulate the following design requirements:

\begin{enumerate}[label=DR\arabic*, ref=DR\arabic*]
\item \label{dr:prioritize} \textbf{Prioritize tracing and evaluation.} In the IDE-native toolset we are building, tracing/observability and evaluation must be high-priority capabilities.
\item \label{dr:datasetless} \textbf{Datasetless start.} Evaluation should naturally incorporate data from manual testing and production signals; it must not require a pre-existing dataset.
\item \label{dr:education} \textbf{Education.} We should educate through evaluation (e.g., through lightweight guidance or read-to-use templates)
\end{enumerate}

\begin{figure*}[t]
  \centering
  \scalebox{1.0}{
  \includegraphics[width=\textwidth,keepaspectratio,clip]{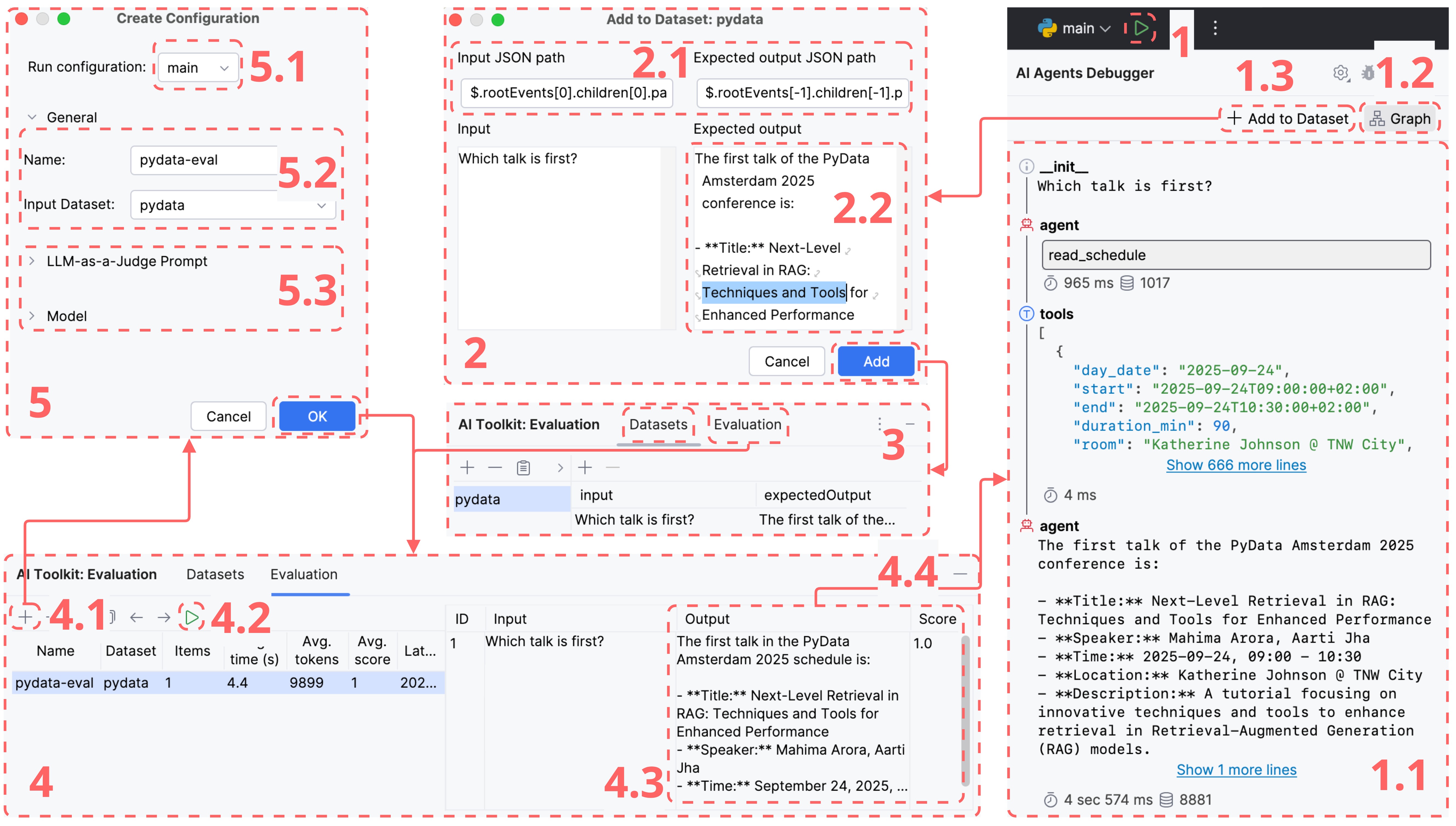}
  }
  \caption{AI Toolkit observe–evaluate flow.}
  \label{fig:ait_flow}
\end{figure*}

\section{Design and Implementation}

We present the AI Toolkit plugin\footnotemark[1] for PyCharm~\cite{pycharm}, consisting of two main components: AI Agents Debugger and AI Evaluation, aligning with \hyperref[dr:prioritize]{\ref*{dr:prioritize}}. The former auto-captures and visualizes traces of LLM/agent launches at Run/Debug; the latter turns examples into datasets and runs repeatable evaluations with pluggable metrics.

Our approach is to surface AI observability and evaluation directly at the developer's \emph{Run} moment within the IDE, turning an ad hoc cycle (“try a prompt, look at the output”) into a tight loop of \emph{trace} $\rightarrow$ \emph{inspect} $\rightarrow$ \emph{curate to dataset} $\rightarrow$ \emph{evaluate} -- with near-zero setup and without switching to external dashboards.

In order to achieve high adoption, we need to make it an ``easy journey'' for the users. Thus, we optimize for three requirements:
\begin{enumerate}[label=UX\arabic*, ref=UX\arabic*]
\item \label{ux:target} \textbf{Precise target action:} Propose value at the exact moment when a user needs it and has time for it.
\item \label{ux:clarity} \textbf{Clarity of value:} Show immediate, visible results.
\item \label{ux:activation} \textbf{Low activation energy:} Require minimum configuration, avoid code edits and external accounts for a useful baseline.
\end{enumerate}

\subsection{Tracing and Observability in the Run Loop}

Modern tracing tools (e.g., LangSmith~\cite{langsmith}, LangFuse~\cite{langfuse}, and Weights \& Biases~\cite{wandb,weave}) capture prompts, tool calls, and intermediate states, but they typically assume outside-the-IDE workflows. We instead bind tracing to the IDE’s \emph{Run} action, so that observability appears precisely when code executes. To cover the broader agentic Python ecosystem, we support the major agentic framework, LangChain~\cite{langchain}, including LangGraph~\cite{langgraph}.

Our tracer uses a client–server architecture. At runtime, a lightweight client, embedded through a Python wrapper, intercepts framework events via callbacks, serializes them, and streams them to an IDE-side server over TCP. The server deserializes, processes, and renders the events in a tool window (\cref{fig:ait_flow}, step~1.1). When available, the client also emits the agent’s execution graph, which the server renders in a dedicated \emph{Graph} tab (\cref{fig:ait_flow}, step~1.2).

Tracing is enabled only for projects that the plugin classifies as \emph{AI projects}. The plugin indexes project files for imports of \texttt{langchain} or \texttt{langgraph}; if any are found, it marks the project accordingly.

When the user clicks \textbf{Run} or \textbf{Debug} (\cref{fig:ait_flow}, step~1), the plugin starts the local server and launches the user’s code under the Python wrapper. The wrapper establishes the TCP channel and attaches a LangChain \texttt{CallbackHandler}~\cite{langchaincallbacks} to each \texttt{Runnable}~\cite{langchainrunnable}. The handler captures events (node entry/exit, LLM and tool calls, and other events), serializes them, and streams them to the server.

The tool window updates incrementally as events arrive, making the tracer compatible with \textbf{Debug} execution: the view reflects the run ``so far,'' enabling step-by-step debugging of agent behavior.

Unlike many existing solutions~\cite{langsmith,langfuse,weave}, we default to a \emph{Pretty} view that surfaces only the signal-bearing fields of each event. In \cref{fig:ait_flow}, for example, the first \texttt{agent} node shows only the invoked tool (\texttt{read\_schedule}), while the second \texttt{agent} node shows only the LLM’s last message. This preserves a readable narrative of the run without the verbosity of full callback payloads.

For each node, the tracer displays execution time and token usage, enabling quick profiling of latency and cost.

This design satisfies \hyperref[ux:target]{\ref*{ux:target}}, \hyperref[ux:clarity]{\ref*{ux:clarity}}, and \hyperref[ux:activation]{\ref*{ux:activation}}. For \hyperref[ux:target]{\ref*{ux:target}}, the tracer appears exactly when the user clicks \textbf{Run}, and it activates only in projects that already depend on LangChain~\cite{langchain} or LangGraph~\cite{langgraph}. For \hyperref[ux:clarity]{\ref*{ux:clarity}}, we show immediate, visible results: the tracer tool window appears automatically, and it updates dynamically during the agent’s run; it also displays the agent’s execution graph, helping visualize the code-defined structure. For \hyperref[ux:activation]{\ref*{ux:activation}}, configuration is minimal -- the plugin wraps the user’s code automatically and wires callbacks to the frameworks -- keeping activation energy low.

\subsection{Evaluation: Unit-tests-like for AI}
To make evaluation habitual for software engineers, we model it after unit tests: developers declare data points as test cases (no hand-written evaluation loops) and run them on the current code version without code changes. In this model, a \emph{dataset} is a table of inputs with reference outputs; an \emph{evaluation} executes the user’s code/agent over each input, collects the generated outputs, and scores them with a configurable set of \emph{evaluators}. Evaluators include standard metrics, LLM-as-a-judge~\cite{llmasjudge}, and lightweight custom code that returns a run score (see step~4 in Figure~\ref{fig:ait_flow}).

From any captured trace in the AI Agents Debugger described earlier, developers can promote the example directly into the dataset: inputs are auto-extracted while reference outputs remain editable, closing the manual-testing loop and providing a low-friction curation path -- thus fulfilling \hyperref[dr:datasetless]{\ref*{dr:datasetless}} (see Figure~\ref{fig:ait_flow}, step~1.3).

AI Toolkit then runs the user’s workflow through any in-IDE run configuration, chosen during the Evaluation Configuration setup (Figure~\ref{fig:ait_flow}, step~5.1). Under the hood, we instrument the Python process at launch -- monkey-patching the relevant I/O boundaries -- to propagate dataset inputs into the system without code edits, aligning with \hyperref[ux:activation]{\ref*{ux:activation}}. The AI Toolkit captures the run’s trace, and we reapply the output JSON path specified during dataset curation (Figure~\ref{fig:ait_flow}, step~2.1) to automatically extract the generated outputs.

The results are presented in a bottom tool window analogous to a unit-test runner (Figure~\ref{fig:ait_flow}, step~4), where each datapoint appears as a test case. Developers can inspect the generated output, evaluator scores, and -- when using LLM-as-a-judge~\cite{llmasjudge} -- the optional explanation. Aggregated statistics such as average score, token usage, and wall-clock time are also summarized.

We persist the complete trace per data point so that when an output is undesired, the developer can look at the internal calls and prompts to understand what happened (Figure~\ref{fig:ait_flow}, step~4.4).

We store datasets and artifacts as project files (YAML files), making them easy to review, diff, and version with standard VCS. This also enables lightweight collaboration by sharing datasets and produced artifacts through version control.

Execution can be offloaded to remote, long-running jobs orchestrated by the Cadence (ex-JetTrain~\cite{jettrain}) backend; progress streams into the IDE with resumable logs and artifact sync, supporting parallelism and recovery without leaving the development environment.

Fulfilling \hyperref[ux:clarity]{\ref*{ux:clarity}} is challenging because users must complete several steps before seeing results (cf. Figure~\ref{fig:ait_flow}). To assist first-time users -- and to meet \hyperref[dr:education]{\ref*{dr:education}} -- we provide a lightweight onboarding flow (not shown in the figure): contextual tooltips, an “Add to dataset” suggestion from a recent trace, prefilled evaluation templates, and a one-click first run that produces an initial score and artifacts.

\subsection{End-to-end Flow Overview}

Rather than pushing developers to external dashboards, we anchor observability and evaluation to the IDE’s \emph{Run} action, compressing the informal “try–look into” routine into an in-IDE loop of \emph{trace} $\rightarrow$ \emph{inspect} $\rightarrow$ \emph{curate to dataset} $\rightarrow$ \emph{evaluate}, with minimal setup and no context switching, as illustrated in Figure~\ref{fig:ait_flow}:

\begin{enumerate}
  \item \textbf{Run \& explore traces.} The user clicks \emph{Run} to execute the agent (1); the AI Agents Debugger shows the trace hierarchy (1.1); optionally, the agent graph is displayed (1.2).
  \item \textbf{Curate dataset from a trace.} User can choose “Add to Dataset” (1.3); then the “Add to Dataset” dialog opens (2); the user configures JSON paths for input and output extraction (2.1) -- the same output path is later reused to automatically extract generated outputs during evaluation -- and verifies or adjusts the expected output (2.2).
  \item The “Evaluation~>~Datasets” view is shown (3), followed by the “Evaluation~>~Evaluation” view (4).
  \item \textbf{Create Evaluation Configuration.} User clicks the “+” button to start creating an Evaluation Configuration (4.1); the “Create Configuration” window opens (5); the user selects an existing agent Run configuration to execute evaluations (5.1), chooses a name and input dataset (5.2), and configures the LLM-as-a-judge~\cite{llmasjudge} prompt and model (5.3).
  \item \textbf{Run Evaluation and inspect.} Back in the Evaluation view (4), the user clicks \emph{Run} to launch the evaluation (4.2); outputs are shown per row (4.3); for any datapoint, the corresponding trace can be inspected in the AI Agents Debugger (4.4).
\end{enumerate}

\section{Release Outcomes}

We have released AI Toolkit as a separate plugin, which one needs to install to start using it. Then we made a promo campaign in PyCharm~\cite{pycharm} for the AI Toolkit following the PyCharm 2025.2 release. At the moment, AI Toolkit only contained AI Agents Debugger functionality with only LangGraph~\cite{langgraph} supported. The promo campaign included (i) the popup appearing after \emph{Run} button was clicked in the projects with LangGraph, suggesting to install the AI Toolkit plugin (ii) description of the plugin in ``What's new`` tab, shown after PyCharm update.

\paragraph{Reach and interest.}
In the four weeks following the PyCharm 2025.2 release, we saw \emph{970} new users of the AI Toolkit. The popup window had a pretty high installation rate (58\%), and generated a one-third of all new users. The other two-thirds arrived from other sources, mainly the ``What's new`` tab.

\paragraph{Activation.}
About \emph{40\%} of those who installed via popup, captured a first trace. However, only around \emph{12\%} of users who installed via ``What's new``, captured a first trace.
The main stopper is the frameworks coverage, not intent: most newcomers (particularly those arriving from “What’s New”) did not use LangGraph in the current project, and the released AI Agents Debugger implementation only supported LangGraph-based agents. Users in unsupported setups typically ignore the plugin rather than churn.

\paragraph{Retention and churn.}
By week four, \emph{26\%} of all installers remained active; among developers who actually made any runs in PyCharm, week-four retention rises to \emph{40\%}. Churn (disable/uninstall) remained low at roughly \emph{10–15\%}.

\paragraph{Takeaways.}
High conversion at \emph{Run}, solid week-four retention, and low churn indicate that the IDE-native, run-triggered workflow is legible and valuable. The principal opportunity is to broaden the framework support (beyond LangGraph).%

\section{Next Steps \& Limitations}
\paragraph{Next steps}
\begin{itemize}
  \item \textbf{Broader framework coverage.} Early indicators suggest that activation will be capped by framework support rather than developer intent. To reach a broader user base, we will release an updated plugin with planned support for the Kotlin agentic framework \emph{Koog}~\cite{koog}, \emph{PydanticAI}~\cite{pydantic, pydanticai} and \emph{OpenAI Python SDK}~\cite{openaisdk}. Following these releases, we will monitor usage signals and metrics to guide further development and extend support to additional agent stacks as demand evolves.

  \item \textbf{Track the usage of AI Evaluation.} We will monitor the adoption of AI Evaluation by collecting usage statistics. Based on feedback, we will prioritize UX improvements and tighter integration between AI Agents Debugger and AI Evaluation.
\end{itemize}

\paragraph{Limitations and threats to validity.}
\begin{itemize}
  \item \textbf{Narrow initial coverage.} The reported activation reflects a release that supported only \emph{LangGraph}, leaving much of the agentic-framework space uncovered; consequently, true demand is likely under-estimated.
  \item \textbf{Short observation window.} Metrics summarize only the first four weeks of a single release, with a sample of roughly 1{,}000 users. A larger sample and a longer observation window will yield more precise estimates.
  \item \textbf{PyCharm-only sample bias.} We have released a plugin only for \emph{PyCharm}~\cite{pycharm}, while a large portion of developers use other IDEs~\cite{so2025-tech-dev-envs-prof-ai}, such as \emph{Visual Studio Code}~\cite{vscode}.
\end{itemize}

\section{Related work}

The rapid development of AI-based applications, particularly those powered by autonomous or semi-autonomous agents, is transforming the software landscape~\cite{gao2024empowering, ning2025survey, sun2024building, jin2024teach}. As these AI agents grow more complex and integrated into diverse workflows, the need for effective tools to monitor, evaluate, and improve their behavior becomes increasingly critical~\cite{chan2024visibility, zheng2025agentsight, dong2024agentops, chan2025infrastructure}.

There are several tools that provide agent observability features at scale. Among them are LangSmith~\cite{langsmith} by LangChain~\cite{annam2025langchain}, Weights \& Biases~\cite{wandb,weave}, MLflow~\cite{mlflow}, and LangFuse~\cite{langfuse}. As mentioned earlier, these capabilities are typically offered as separate tools or dashboards, designed for ML practitioners who already know which controls to use and how to integrate them effectively.

Evaluation of AI agents is an active area of research~\cite{zhugeagent, yin2025mmau, yehudai2025survey} and industry~\cite{arizeai, Ip_deepeval_2025, superannotate}.
Research efforts focus on developing general evaluation approaches (e.g., human-based, LLM-as-a-judge~\cite{llmasjudge}), assessing agent capabilities (e.g., planning, function calling), and designing evaluation methods for particular domains (e.g., software engineering agents, data agents)~\cite{yehudai2025survey}.
Industrial solutions emphasize collaboration features and integration with existing customer infrastructure.
The main goal of this work is to provide a simple, usable evaluation flow directly within the IDE.

A related initiative is the VS Code AI Toolkit by Microsoft~\cite{vscode-ai-toolkit}. It follows a similar goal of improving the developer experience for AI workflows but focuses on a different set of capabilities. Specifically, it offers tight integration with Azure AI Studio for model fine-tuning, deployment, and access to a model catalogue, while also including basic evaluation and traceability features. In contrast to our approach, which emphasizes lightweight, IDE-native evaluation and observability, the VSCode AI Toolkit is designed as part of a broader Azure ecosystem that streamlines the connection between local development and cloud-based model management.

\section{Conclusion}
By meeting developers in the IDE and guiding tracing, evaluation, and experimentation as a single workflow, the AI Toolkit lowers cognitive overhead and makes disciplined AI development accessible to non-ML specialists. This paper contributes: (i) empirical evidence of developer needs around non-determinism and ad hoc testing; (ii) an IDE-native design that ties observability and evaluation to the Run/Debug moment, including run-triggered tracing and a curated path from trace $\rightarrow$ dataset $\rightarrow$ evaluation; and (iii) an implementation in JetBrains IDEs (AI Agents Debugger and AI Evaluation) that delivers these capabilities with near-zero setup.

Early usage indicates that this approach is legible and valuable: developers respond to in-context prompts at Run, successfully capture traces, and retain the tool at meaningful rates; the main activation bottleneck is framework coverage rather than intent. While results reflect a short window and an initially narrow scope, they provide actionable direction.

More broadly, we argue that bringing AI observability and evaluation into the IDE -- rather than exporting developers to external dashboards -- offers a practical path to AI features in real projects. We invite the IDE and SE communities to build on this pattern: standardize trace schemas, treat evaluations as first-class tests, and connect IDE-native feedback loops to CI and production telemetry.

\begin{acks}
We thank Mikhail Bogdanov from the Survey team of the Strategic Research \& Market Intelligence department of JetBrains for designing, collecting, cleaning, and analyzing survey data, and Olga Galchenko, Artem Sarkisov, and Sofia Kulikova for their initiative, for contributing to questionnaire development, and for their domain expertise.
\end{acks}

\bibliographystyle{ACM-Reference-Format}
\balance
\bibliography{ai-toolkit}

\end{document}